\documentclass{article}
\usepackage{amsmath}
\usepackage{epsfig}\parskip5pt 
\begin{document}\numberwithin{equation}{section}

\title{ Theory of small aspect ratio waves in deep water}
\author {R.A. Kraenkel \ddag, J.Leon \dag and M. A. Manna \dag\\
 \dag {\em Physique Th\'eorique et Astroparticules, CNRS-UMR5207,}\\
Universit\'e  Montpellier II, 34095 Montpellier (France)\\
 \ddag {\em Instituto de F\'{\i}sica Te\'orica-UNESP,}\\
Rua Pamplona 145, 01405-900 S\~ao Paulo (Brazil)}
\date{}\maketitle

\begin{abstract} In the limit of small values of the aspect ratio parameter
(or wave steepness) which measures the amplitude of a surface wave in units of
its wave-length, a model equation is derived from the Euler system in infinite
depth (deep water) without potential flow assumption. The resulting equation is
shown to sustain periodic waves which on the one side tend to the proper linear
limit at small amplitudes, on the other side possess a threshold amplitude
where wave crest peaking is achieved. An explicit expression of the crest angle
at wave breaking is found in terms of the wave velocity. By numerical
simulations, stable soliton-like solutions  (experiencing elastic
interactions)  propagate in a given velocities range on the edge of which they
tend to the peakon solution.  
\end{abstract}

{\em Published : Physica D 211 (2005) 377-390}

\section{Introduction}

The description of the propagation of surface waves in an ideal incompressible
fluid is still a classical subjet of investigation in mathematical physics as
no definite comprehensive answer to the problem has been given yet. In the
limit of {\it shallow water}, surface gravity waves have been intensively
studied and many model equations were introduced by various approaches, with
great success. The {\it nonlinear deep water} case is more cumbersome and there
does not exist today a simple model as universal as the  shallow water 
equations (Korteweg-de Vries or Boussinesq) which would result from an
asymptotic limit of the Euler system.

The inherent technical differences between shallow and deep water are mainly
due to the fact that the two {\it natural} small parameters used for
perturbative analysis of the Euler system in shallow water loose their sense in
the deep water case (depth $h\to\infty$). Indeed, these parameter are
$\epsilon_1=a/h $, which measures the amplitude $a$ of the perturbation scaled
to the fluid depth $h$, and $\epsilon_2=h^2/\lambda^2 $, which measures the
depth in units of wavelength $\lambda$.

By perturbative  expansion in $\epsilon_1$, and fixed finite $\epsilon_2$, one
obtains the nonlinear shallow water equation \cite{whitham}. Retaining
$\epsilon_1$ and $\epsilon_2$ (but not their product) leads to different
versions of the Boussinesq  equations \cite{boussinesq}, from which the 
Korteweg-de Vries equation  is derived by assuming a small amplitude wave
moving in a given direction  \cite{korteweg}.  All these model equations govern
asymptotic dynamics of  long wavelength {\it wave profiles}.

One way to obtain a small parameter in deep water is to take into account that
deep water waves typically result from a superposition of wave components more
or less close to a fundamental carrier wave.  The small parameter then measures
the variations of the envelope scaled to the one for the carrier wave.  Such
{\it nonlinear modulation of wave trains} is worked out perturbatively by means
of {\it slowly varying envelope approximation} (SVEA) which usually leads in
1+1 dimensions to the nonlinear Schr\"odinger model \cite{benney,zakh}. For a
full account on modulation of short wave  trains on water of intermediate or
great depth we refer to   \cite{mei,dias}. The procedure provides  nonlinear
dynamics of surface waves under the form of nonlinear modulation, and the
drawback is that  the dynamics of the wave profile itself remains unknown. 

Based on  the theory of analytical functions and perturbation theory, a model
for the profile of the free surface wave in water of finite depth, involving
the Hilbert transform operator, was derived in \cite{matsuno}.  Although this
model possess a well-defined deep water limit, the resulting equation cannot be
studied by known techniques to compare it to KdV-like models. Other model
equations, built to fit the properties of waves on deep water can be found in
\cite{whitham2,bro}. Their dispersion relations coincide exactly with that of
the water waves on infinite depth but their nonlinear terms are chosen {\it ad
hoc} to reproduce Stokes waves.

Our purpose here is to study the asymptotic dynamics of the very profile of a
surface wave in deep water in the weak nonlinear limit by assuming  a 
dependence on the vertical coordinate close to the linear harmonic one. The
dispersion relation of a fluid on a depth $h$
\begin{equation}\label{gener-disp}
\omega^2=gk\tanh(kh),\end{equation}
leads for long waves on shallow water (parameter $kh$ small) to the 
nondispersive relation
\begin{equation}
\omega=k\sqrt{gh}\ \Rightarrow \ v_p=v_g= \sqrt{gh},\end{equation}
where $v_p$ is the phase velocity and $v_g$ the group velocity. However,
waves on deep water (parameter $kh\to\infty$) are dispersive as from
\eqref{gener-disp}
\begin{equation}\label{deep-disp}
\omega=\sqrt{gk}\ \Rightarrow \ v_p=\sqrt{\frac gk},\quad 
v_g=\frac12\sqrt{\frac gk}.\end{equation}
This {\it deep water dispersion relation} will be one of our main guide in
the process of finding a limit model whose linear limit is constrained by 
\eqref{deep-disp}.

Our approach follows the method of Green and Naghdi  \cite{GLN1,GLN2} 
for surface waves in shallow water which assumes an anzatz for the dependence
of the velocity components on the vertical dimension $z$. This anzatz does not
produce an exact solution of the full Euler system and the game consists in
replacing one of the equations with its {\it integrated expression}. This comes
actually to making an average over the depth, which can be performed with
different weights. Although weight is not determinant in the shallow water case
\cite{johnson}, we shall see that its choice is prescibed by consistency
requirement at the linear limit.

A limit model is then obtained by defining a small parameter which measures the
amplitude of a surface wave in units of its wave-length, we call it the  {\it
aspect ratio parameter} (ARP). The approach combines the asymptotic analysis
{\it \`a la Whitham} with the already standard method of  multiple scales
\cite{tan,pan,can,pereira}. This will be shown to lead to the following model
\begin{equation}\label{main}
\eta_t - \eta_{xxt} - \frac12\eta_{xxx} + \frac32 \eta_x + 
\eta\eta_x = \frac5{3}\eta_x\eta_{xx} +\frac1{3} \eta\eta_{xxx}.
\end{equation}
for the dimensionless deformation $\eta(x,t)$ of the free surface of
deep water.

The paper is organized as follows. In section \ref{euler} we introduce the
Euler equations, their nondimensional version and the anzatz which, together
with convenient average, enables to reduce the initial three-dimensional
problem to a two-dimensional one. The resulting model, equations \eqref{GN} 
below, is in fact the deep water version of the celebrated Green-Naghdi  system
for surface waves in shallow water \cite{GLN1,GLN2}.  

In section \ref{asymptotic}, the nonlinear dispersive model \eqref{main}  that
governs propagation of waves of small ARP in deep water is derived by the
method of multiple scales and perturbative expansion. Expressed in physical
units, the model appears with $k$-dependent coefficients, reminiscent of what
occurs in the SVEA approach of deep water modulation. 

In section \ref{stokes}, analytical and  numerical analysis of the progressive
periodic waves are performed. Numerical simulations show the existence of
periodic waves than tend to peak as amplitude grows. The crest angle at the peak
is calculated and we obtain an explicit expression in terms of the wave
velocity. Moreover numerical simulations show also the existence of soliton-like
solutions (though we cannot provide analytic expression) that tend, at the edge
of the allowed  velocity range, to the peakon solution (who is given an explicit
analytical expression).

\section{The model equations in deep water} \label{euler} 

\subsection{General settings.}

Let us consider the Euler equations in physical dimensions where the particles
of the fluid are identified in a fixed rectangular  Cartesian system of center
$O$ and axes $(x, y, z)$ with  $Oz$ the  upward vertical direction.  We assume
translational symmetry in $y$  and thus consider a sheet of fluid in the $xz$
plane. The velocity of the fluid in this plane is the vector $(U,W)$ where
$U(x,z,t)$ is the horizontal component and $W(x,z,t)$ the vertical one.  This
fluid sheet is moving on a  bottom at $z=-\infty$ and its free surface is
$z=\eta(x,t)$. 

The continuity equations and the Newton equations in the flow domain read
\begin{align} 
U_x + W_z &= 0 ,\label{cont} \\
\rho \dot{U} &= -P _{x}   ,  \label{newx}\\
\rho \dot{W} &= -P _{z} -g\rho , \label{newz}
\end{align}
where $P (x,z,t)$ denotes the pressure, $\rho$ the uniform density and $g$ is
the gravity. Subindices mean partial derivatives and overdot means  material
derivative defined as usual by
\begin{equation}\label{matder}
\dot{F} =  F_t + UF_x + WF_z.
\end{equation}
The boundary conditions at $z = -\infty$ simply state that the velocity
component vanish ($U \to0$ and $W\to0$), while boundary conditions at the free
surface $z =\eta (x,t)$ state that $P $ is the atmospheric pressure $P_a$ and
that the total derivative of the surface equation $z-\eta=0$ vanishes, namely
\begin{equation}
 z = \eta \ : \quad   \eta _{t}+U \eta_{x}-W=0.
\end{equation}
We are interested here in finding the evolution equations for the free surface
by studying the {\it nonlinear deformation} of a particular linear wave profile
with given arbitrary wave number $k$.

\subsection{Dimensionless Euler system.}

The linear  progressive monochromatic solution of the linear limit of 
the above Euler system reads \cite{light}
\begin{equation}\label{Ulinear}
U_{\rm linear}=U_0\cos (kx - \omega t)\ e^{kz},\quad \omega^2 = gk,
\end{equation}
where the dispersion relation is indeed of the deep-water class 
\eqref{deep-disp}.

Thus, given the wave number $k$, we may scale the  original space variables 
$x$ and $z$ with $k$, the time variable with $\sqrt{kg}$, the velocity
components $U$ and $W$ with $\sqrt{k/g}$, and the pressure $P $ with  $k/\rho
g$. For convenience we keep the same notations for the adimensional variables
and the Euler equations for $z\in[-\infty,\eta]$ then become
\begin{align} 
U_{x} + W_{z} &= 0, \label{cont2}\\
\dot{U} &= -P _{x} , \label{newx2}\\
\dot{W} &= -P _{z} -1 ,\label{newz2}
\end{align}
whith the boundary conditions  
\begin{align}
z \to-\infty\ : \quad & U = W= 0 ,  \label{BC1}\\
z = \eta\ : \quad  & P =P_0 ,\label{BC2}\\
 z = \eta \ : \quad  & \eta _{t}= W- U \eta_{x},\label{BC3}
\end{align}
where  $P_0=P_ak/\rho g$.  Note that, as $z$ is scaled with $k$,  the
dimensionless surface profile  $\eta$ is also scaled with $k$.

\subsection{Vertical velocity profile.}

Inspired thus by \eqref{Ulinear}, we assume an exponential vertical dependence
of the horizontal velocity component and seek an approximate solution of the
Euler system by starting with the anzatz
\begin{equation}\label{anzatz1}
U(x,z,t) = u(x,t)\ e^{z}.
\end{equation}
It does satisfy the boundary condition (\ref{BC1}) and gives
\begin{equation}
W(x,z,t) = -u_x(x,t)\ e^{z},
\label{anzatz2}
\end{equation}
obtained from the continuity equation (\ref{cont2}) and the boundary 
condition (\ref{BC1}).

By computing the total derivative of the above velocity components we get
\begin{align}
&\dot{U} = u_t\ e^z,\label{upunto}\\
&\dot{W} = -u_{xt}\ e^z -uu_{xx}\ e^{2z} +
u_{x}^2\ e^{2z},\label{wpunto} 
\end{align} 
which allows to calculate explicitely the pressure $P $ out of 
(\ref{newz2}) and boundary condition \eqref{BC2} as
\begin{equation}\label{newz3}
  P -P_0=(\eta-z)-u_{xt} (e^{\eta}-e^z)+\frac12 (u_x^2-uu_{xx})
  (e^{2\eta}-e^{2z}).
\end{equation} 
This pressure diverges as $z\to-\infty$ in {\it Archimedean way} as it must.

The boundary condition \eqref{BC3} finally provides the evolution of $\eta$ as
\begin{equation}\label{GN1}
\eta_t + (u\ e^\eta)_x = 0.
\end{equation} 
To that point the anzatzs \eqref{anzatz1} and \eqref{anzatz2}, the  expression
\eqref{newz3} of $P $ and the evolution \eqref{GN1} of the free surface $\eta$
satisfy exactly the differential equations \eqref{cont2} and \eqref{newz2}
together with the whole boundary conditions \eqref{BC1}, \eqref{BC2} and
\eqref{BC3}. The remaining equation to take into account is then the Newton's
law \eqref{newx2} which  is of course not satisfied globally by the above
expressions of $W$ and $P$. 

\subsection{Averaging Newton's law.}

The Newton's law \eqref{newx2} is now taken into account through an average on
the full depth with the weight $e^{\alpha z}$ which regularizes the diverging
Archimedean term in the expression of the pressure \eqref{newz3}. This is
explicitely 
\begin{equation}
\int_{-\infty}^\eta  dz\ e^ {\alpha z}(\dot U+P _x)=0. \end{equation}
which eventually furnishes with use of \eqref{upunto} and  \eqref{newz3}
\begin{equation}\label{GN-alpha}
\frac{e^{(\alpha+1)\eta}}{\alpha+1} u_t =
\frac1{\alpha}
\frac{\partial}{\partial x}\left(\frac{e^{(\alpha+1)\eta}}{\alpha+1} u_{xt}+ 
\frac {e^{(\alpha+2)\eta} } {(\alpha+2)}[uu_{xx}-u_x^2]
-\frac{e^{\alpha\eta}}{\alpha}\right).
\end{equation}
It is a {\it model-dependent} system where the solutions now depend on the
external parameter $\alpha$.

The value of parameter $\alpha$ is fixed by demanding that the linear limit of
the system \eqref{GN1}\eqref{GN-alpha} possess solution with phase and group
velocities given by \eqref{deep-disp}. In dimensionless units, these velocities
are then required to be
\begin{equation}\label{demand}
v_p=1\ ,\quad v_g=\frac12.\end{equation}
In the linear limit of system \eqref{GN1}\eqref{GN-alpha},  $\eta$ can 
easily be eliminated to arrive at
\begin{equation}
\alpha u_{tt}-(\alpha+1)u_{xx}-u_{xxtt}=0,\end{equation}
which possess the solution
\begin{equation}
u=u_0\cos(qx-\omega t),\quad
\omega^2(q)=\frac {(\alpha +1)q^2}{\alpha+q^2}.
\end{equation}
Requiring relation \eqref{demand} for the above dispersion law $\omega(q)$
comes to require the relations
\begin{equation}\label{require}
v_p=\left.\frac {\omega(q)}q \right|_{q=1}=1,\quad 
v_g=\left.\frac {\partial\omega}{\partial q}\right|_{q=1}=\frac12.
\end{equation}
The first of these equations is verified for any $\alpha$ while the second 
holds if and only if $\alpha=1$. 

As a result the basic system of equation is \eqref{GN1} and \eqref{GN-alpha}
with $\alpha=1$ namely
\begin{align}\label{GN}
&\frac12 u_t\ e^{2\eta}+\partial_x\left(e^{\eta}-
\frac12 u_{xt}\ e^{2\eta} + \frac13 e^{3\eta}[u_x^2-uu_{xx}]\right)=0,
\nonumber\\
&\eta_t + (u\ e^\eta)_x = 0.
\end{align}
The system \eqref{GN} for the couple of variables $u(x,t)$ and $\eta(x,t)$ is
the net result of assuming the $z$-dependence \eqref{anzatz1} in the  Euler
equations and of making a convenient average of Newton's law on the depth
$(-\infty,\eta]$.  As displayed above, the crucial step is averaging 
\eqref{newx2} with the precise weight  $e^z$. Indeed, while it can be proved
from \cite{johnson} that weighting the average is not determinant in the
shallow water case, weight is essential here and the choice  $e^z$  is
justified by requirement \eqref{deep-disp}.

\section{Small aspect ratio waves in deep water}
\label{asymptotic}

\subsection{Generalities.}

The small parameter of the perturbative analysis of the system
\eqref{GN} is, in dimensionless units, the maximum amplitude
$\epsilon$ of the free surface deformation $\eta(x,t)$. So we define
\begin{equation}\label{surface}
\eta(x,t) = \epsilon H(x,t), \end{equation}
and it is instructive to understand the physical meaning of $\epsilon$ by
returning to the physical units for which
\begin{equation}\label{stepp} \epsilon = ak ,
\end{equation}
where $a$ is the wave profile maximum amplitude and $k$ the chosen wave number.
Thus $\epsilon$ is the parameter that measures the ratio of wave height to wave
length, we call it the aspect ratio parameter (ARP). Then a small ARP
($\epsilon<1$) means a {\it flat deformation} of the surface without reference
to the absolute amplitude or to the depth.

The dimensionless linear dispersion relation ($\omega^2 = 1$)  undergoes
deviations due to nonlinearity (Stokes' hypothesis) which can be taken into
account through the expansion
\begin{equation}\label{omegapertur}
\omega^2 = 1 + \epsilon + \epsilon^2 + \cdots .
\end{equation}
Therefore the phase expands as
\begin{equation}
x -\omega t =  x - t- \epsilon t - \epsilon^2 t + \cdots ,
\end{equation}
which suggests the following definition of the new variables $y$, $\tau$, 
$\nu$, $\cdots$,
\begin{equation} 
y = x - t, \quad \tau = \epsilon t, \quad \nu = \epsilon^2 t, \quad \cdots .
\end{equation}
The function $ H(x,t) $ is then {\it represented} as the function 
$H(y, \tau, \nu, ...,)$ of the new
variable $y$ and the slow variables $\tau$, $\nu$, $\cdots$, for which hold 
the derivation rules 
\begin{align} 
\partial _x =&\partial_y,
\label{operators1}\\
\partial_t =& -\partial_y +
\epsilon \partial_{\tau} +\epsilon^2 \partial_{\nu} +\cdots .
\label{operators2}
 \end{align}
These rules are now explicitely applied step by step to our system \eqref{GN}.

\subsection{Asymptotic expansion.}

Since $\exp(\epsilon H) = 1 + \epsilon H + O(\epsilon^2)$ we can consider
system \eqref{GN} at order $\epsilon^0$, next at orders 
$\epsilon^0$ and $\epsilon$ and so on. In terms of $H$,
$y$ and $\tau$ we get
\begin{equation}
(-\partial_y + \epsilon \partial_{\tau}+ {\cal O}(\epsilon^2))\epsilon H + [u(1
+ \epsilon H + {\cal O}(\epsilon^2))]_y = 0 
\end{equation}
This equation gives $u$ in term of derivatives and integrations of $H$
\begin{equation}\label{u-H}
u = \epsilon H -\epsilon^2 (H^2 + \int^{y}_{-\infty} H_{\tau}dy') + {\cal O}(\epsilon^3)
\end{equation}
where it is not assumed that $u$ or $H$ (and derivatives) vanish
as $y \rightarrow -\infty$. Instead the above equation readily gives 
the usefull relation
\begin{equation}\label{consintegr}
u(-\infty,\tau) = \epsilon H(-\infty,\tau)[1 - \epsilon H(-\infty,\tau) + 
{\cal O}(\epsilon^2)]. 
\end{equation}
Now equation (\ref{GN1}) is more conveneintly written in the form
\begin{equation}
\frac12 A + \frac13 B- C -\frac12 D = 0
\label{eqABCD}
\end{equation}
with $A, B, C $ and $ D $ given by
\begin{equation}
A = (u_{xt} e^{2\eta})_x, \quad B = ((uu_{xx} - (u_x)^2)e^{3\eta})_x, \quad 
C = (e^{\eta})_x ,\quad  D = u_t e^{2\eta}.
\end{equation}
Using the operator expansion
\begin{equation}
\partial^2_{xt} =  -\partial^2_{yy} + \epsilon \partial^2_{y\tau} + 
{\cal O}(\epsilon^2)
\end{equation}
we obtain
\begin{align*}
A = \frac{\partial}{\partial y}
&\{(1 + 2 \epsilon H + {\cal O}(\epsilon^2))\times\\
 &(-\partial^2_{yy}+\epsilon \partial^2_{y\tau}+{\cal O}(\epsilon^2))
(\epsilon H -\epsilon^2 H^2 \epsilon^2\int^{y}_{-\infty} H_{\tau}dy' + 
{\cal O}(\epsilon^3))\} ,
\end{align*}
which at orders $\epsilon$ and $\epsilon^2$ provides
\begin{equation}\label{eqA}
A = -\epsilon H_{yyy} + \epsilon^2 ( 4H_yH_{yy} + 2H_{yy\tau}) + {\cal O}
(\epsilon^3).\end{equation}

Coming now to the computation of $B$ we may write
\begin{align*}
B = &\frac{\partial}{\partial y}
\{ [1 + 3 \epsilon H + {\cal O}(\epsilon^2)]\times \\
 &[ (\epsilon H -\epsilon^2 H^2 -\epsilon^2 \int^{y}_{-\infty} H_{\tau}dy' + 
{\cal O}(\epsilon^3))(\epsilon H_{yy} +  {\cal O}(\epsilon^2))
- \epsilon^2 (H_ {y})^2 + {\cal O}(\epsilon^3)]\}
\end{align*}
which at order $\epsilon^2$ provides
\begin{equation}\label{eqB}
B = \epsilon^2 (HH_{yyy} - H_yH_ {yy}) + {\cal O}(\epsilon^3).
\end{equation}

Finally $C$ and $D$ at order $\epsilon^2$ are given by
\begin{align}
C = &(e^{\epsilon H})_y = \epsilon H_y  + \epsilon^2 HH_y + 
  {\cal O}(\epsilon^3),\label{eqC}\\
D = &[-\partial_y +\epsilon \partial_{\tau} + {\cal O}(\epsilon^2)]
[\epsilon H -\epsilon^2 (H^2 + \int^{y}_{-\infty} H_{\tau}dy') + {\cal O}
(\epsilon^3)]\times  \nonumber\\
& [1 + 2 \epsilon H + {\cal O}(\epsilon^2)] \nonumber\\
=& -\epsilon H_y + \epsilon^2 2H_{\tau} + {\cal O}(\epsilon^3). 
\end{align}

We substitute $A$, $B$, $C$ and $D$ in (\ref{eqABCD}), keep the 
terms in $\epsilon^0$ and $\epsilon^1$ to obtain eventually 
\begin{equation}
-H_y -H_{yyy} + \epsilon (\frac{10}{3}H_yH_{yy} + \frac23 HH_{yyy} - 2HH_y
+2H_{yy\tau} - 2H_{\tau}) = 0
\label{maininytau}
\end{equation}
or in the original (dimensionless) $x$ and $t$ variables
\begin{equation}
H_t + \frac32 H_x - H_{xxt} - \frac12 H_{xxx} + \epsilon HH_x = \epsilon 
(\frac53 H_xH_{xx} + \frac13 HH_{xxx}).
\end{equation}
In terms of the free surface $\eta=\epsilon H$ the above equation reads
\begin{equation}\label{eta}
\eta_t - \eta_{xxt} - \frac12\eta_{xxx} + \frac32 \eta_x + 
\eta\eta_x = \frac53\eta_x\eta_{xx} +\frac1{3} \eta\eta_{xxx},
\end{equation}
as announced in the introduction. This is the equation  which describes
asymptotic nonlinear  and dispersive  evolution of  small ARP waves in deep
water.

\subsection{Discussion.}

We note first that the evolution \eqref{eta} of the free surface,
written in physical dimensions, namely by $\eta\to k\eta$, $x\to kx$
and $t\to t\sqrt{gk}$, reads 
\begin{align}\label{eqeta}
\eta_t-\frac{1}{k^2}\ \eta_{xxt}  
- \frac{1}{2k^2}\sqrt{\frac{g}{k}}\ \eta_{xxx} 
+ &\frac32 \sqrt{\frac{g}{k}}\ \eta_x 
+\sqrt{gk}\ \eta \eta_x =\nonumber\\
 &\frac1{3k} \sqrt{\frac{g}{k}}\ (5\eta_x \eta_{xx} +  \eta \eta_{xxx}) .
\end{align}
It is therefore a $k$-dependent equation that describes the nonlinear
deformations of the wave of given $k$.

Next, by  defining
\begin{equation} 
F(x,t) = \frac13\,\eta(x,t) + \frac12
\end{equation}
we transform the equation (\ref{eta}) into 
\begin{equation}\label{CHtype}
F_t - F_{xxt} + 3FF_x = 5F_xF_{xx} + FF_{xxx}.
\end{equation}
Equations of the form
\begin{equation}
F_t - F_{xxt} + (b + 1)FF_x= bF_xF_{xx} + FF_{xxx},
\end{equation}
are completely integrable only in the cases $b = 2$  (Camassa-Holm equation)
\cite{ch} and $b = 3$ (Degasperis-Procesi equation) 
\cite{Procesi,Degas,Dullin}.  Our equation (\ref{CHtype}) does not enter this
class and we cannot tell more about integrability.

Finally, from the above perturbative expansion we also have, according to
\eqref{u-H} the same equation for the variable $u(x,t)$, which can also be
written as the conservation law
\begin{equation}\label{u}
\partial_t(u - u_{xx})=\partial_x(\frac12u_{xx} -\frac32 u -
\frac12u^2 +\frac23u_x^2 +\frac1{3} uu_{xx}).
\end{equation}

\section{Progressive waves}
\label{stokes}

\subsection{Periodic progressive waves.}

The waves of constant profile propagating at velocity $c$ are the solutions of 
\eqref{eta} for
\begin{equation}
\eta(x,t)=\eta(\xi)\ ,\quad \xi=x-ct,\end{equation}
which implies the ordinary nonlinear differential equation
\begin{equation}\label{eta-stat}
(\frac32-c)\eta_{\xi}- (\frac12-c)\eta_{\xi\xi\xi} + 
\eta\eta_{\xi}= \frac5{3}\eta_{\xi}\eta_{\xi\xi} +
\frac1{3} \eta\eta_{\xi\xi\xi},
\end{equation}
which can be integrated once to give
\begin{equation}\label{eta-integ}
(\frac32-c)\eta- (\frac12-c)\eta_{\xi\xi} + 
\frac12\eta^2 -\frac23\eta_{\xi}^2-\frac13 \eta\eta_{\xi\xi}=A,
\end{equation}
where $A$ is an integration constant to be determined from the chosen
boundary conditions.

\begin{figure}[ht]
\centerline{\epsfig{file=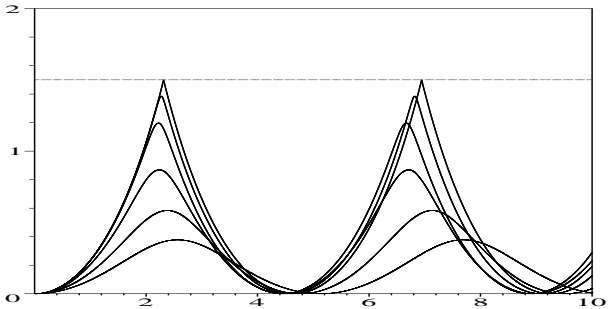,height=4cm,width=8cm}}
\caption {Profiles of the dimensionless surface deformation $\eta$ solution of
\eqref{eta-stat} with \eqref{init} for $c=1$ and $\alpha$ taking the values
$0.2$, $0.3$, $0.4$, $0.446$, $0.4499$ and $0.45$, from bottom to top.}
\label{fig:profile1}\end{figure}

To study the periodic solutions, we integrate \eqref{eta-stat}  numerically 
under the initial values
\begin{equation}\label{init}
\eta(0)=0,\quad \eta_\xi(0)=0,\quad \eta_{\xi\xi}(0)=\alpha,\end{equation}
and obtain a typical set of profiles displayed in figure \ref{fig:profile1}
for different values of the parameter $\alpha$.
Such a simulation shows the wave crest peaking, together with wavelength
variations, as wave amplitude increases. We have checked that low amplitude
profiles (obtained for small values of $\alpha$) do tend to the linear
solution $\alpha(\cos(\xi)+1)$.

The value of the maximum amplitude $\eta_s$ reached at wave peaking can be
infered from the equation \eqref{eta-integ}, it is the value for which the 
coefficient of the second derivative vanishes, namely  
\begin{equation}\label{threshold}
\eta_s=3\ (c-\frac12).\end{equation}
We have found actually that this occurs for the initial data  \eqref{init} with
the threshold $\alpha_s$  given by
\begin{equation}\label{init-threshold}
\alpha_s=\frac3{10}\  (\frac52-c).\end{equation}
At the above threshold values, the crest angle can be evaluated as follows.

\subsection{Crest angle.}

 First the integration constant $A$ of  \eqref{eta-integ} is evaluated at
$\xi=0$ by means of \eqref{init} and \eqref{init-threshold},  we obtain
\begin{equation}
A=\frac3{10}(c-\frac12)(\frac52-c).\end{equation}
Next we evaluate equation \eqref{eta-stat} at the limit $\xi\to \xi_s^\pm$ 
(limit to the left $\xi_s^-$ or to the right $\xi_s^+$), where $\xi_s$ is the
value  for which the threshold is reached, namely $\eta(\xi_s)=\eta_s$.  We
obtain this way the limit of the second derivative by
\begin{equation}
\lim_ {\xi\to \xi_s^\pm}(\eta_{\xi\xi})=\frac65v.\end{equation}
\begin{figure}[ht]
\centerline{\epsfig{file=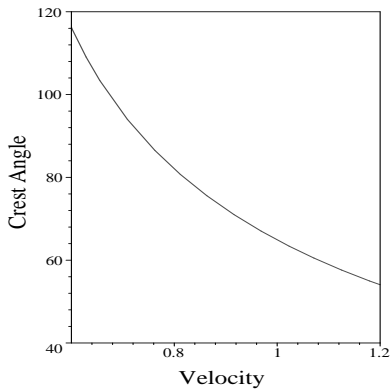,height=5cm,width=5cm}}
\caption {Values of the crest angle, in degrees, at wave peaking for velocities
$c$ in the range $[0.6, \ 1.2]$.}
\label{fig:angles}\end{figure}

To get the limits of the first derivative we evaluate now equation
\eqref{eta-integ} at $\xi=\xi_s^\pm$. With help of the above value of $A$ we
obtain
\begin{equation}
\lim_ {\xi\to \xi_s^\pm}(\eta_{\xi}^2)=
 \frac94 \; (c-\frac12)\;(\frac65c+1),
\end{equation}
and thus the limit of $\eta_{\xi}$ in $\xi_s^\pm$ can be either positive or
negative. It is then clear that a bounded periodic solution like the one
displayed on figure \ref{fig:profile1} implies the following solution
(remember $c>1/2$)
\begin{equation}\label{slope}
\lim_ {\xi\to \xi_s^-}(\eta_{\xi})=-\lim_ {\xi\to \xi_s^+}(\eta_{\xi})=
 \left[ \frac94 \; (c-\frac12)\;(\frac65c+1)\right]^{1/2}.
\end{equation}
Last, from the above value of the slope at $\xi_s$, the crest angle at wave 
peaking reads
\begin{equation}\label{angle}
\theta=2\ \arctan\left\{ 
\left[ \frac94 \; (c-\frac12)\;(\frac65c+1)\right]^{-1/2}\right\},
\end{equation}
and the graph of this expression is plotted on the figure \ref{fig:angles}.
At $c=1$ the angle is $\theta=65\ d^o$.

It is worth remarking that the analytic expression of the crest limit angle,
which we have cheked on numerical simulations, actually reults in a part from
the phenomenological expression \eqref{init-threshold} obtained by observations
of numerical simulations of equation \eqref{eta-stat} with boundary data
\eqref{init}. 

\subsection{Peakon solution.}

The coefficients of the ODE  \eqref{eta-stat} show that periodic solutions 
hold if $c>1/2$ and a threshold is reached at
$c=5/2$ for which waves of vanishing amplitudes are already singular. At this
value, the model supports the  {\em peakon solution} 
\begin{equation}\label{peakon}
\eta (\xi)=6\exp [-|\xi|/\sqrt{2}],\quad \xi=x-\frac52t\ .\end{equation}
Remarkably enough, this peakon also appears as the limit of the solitary wave
solution when its velocity reaches the value $5/2$.

\subsection{Soliton-like waves.}

Numerical simulations has revealed the existence of stable soliton-like
excitations that travel without deformation at constant speed in the range
$[3/2,\ 5/2]$. These are performed by observing that
\begin{equation}\label{soliton}
\eta^{(s)}=\frac {\eta_0}{\cosh^2(\kappa(x-vt-x_0))},
\quad \eta_0=3\ (v-\frac32),\quad
\kappa=\frac12\sqrt {\frac{2v-3}{2v-1}},
\end{equation}
is an approximate solitary wave solution for \eqref{main} according to
\begin{equation}
\eta_t^{(s)} - \eta_{xxt}^{(s)} - \frac12\eta_{xxx}^{(s)} + 
\frac32 \eta_x^{(s)}+\eta^{(s)}\eta_x^{(s)}-\frac5{3}\eta_x^{(s)}\eta_{xx}^{(s)}
-\frac1{3}\eta^{(s)}\eta_{xxx}^{(s)}
={\cal O}(\kappa^3\eta_0^2).
\end{equation}
Hence defining
\begin{equation}
v=\frac32+\delta\ \Rightarrow\ \{\eta_0=3\delta,
\ 4\kappa^2=\frac \delta {1+\delta}\},\end{equation}
expression \eqref{soliton} tends to the exact solution for $\delta\to0$
(for which the amplitude $\eta_0$ vanishes).

We have performed a series of numerical simulations of \eqref{main} 
by first writing it in the frame moving at velocity $v$, namely
\begin{equation}
\eta_t+(\frac32-v)\eta_x - \eta_{xxt} +(v-\frac12)\eta_{xxx} + 
\eta\eta_x=\frac5{3}\eta_x\eta_{xx}+\frac1{3}\eta\eta_{xxx}.
\end{equation}
and then by using the following initial-boundary value problem on $x\in[0,L]$
\begin{equation}
\eta(0,t)=\eta_s(x,0),\quad
\eta(0,t)=0,\quad \eta(L,t)=0,\quad \eta_x(L,t)=0,\end{equation}
with an initial position $x_0=2L/3$ and a velocity
\begin{equation}
v=\frac32+\delta,\quad \delta\in]0,1[.
\end{equation}
The method uses a standard second
order implicit finite difference scheme with a mesh grid dimension $0.1$.

\begin{figure}[ht]
\centerline {\epsfig{file=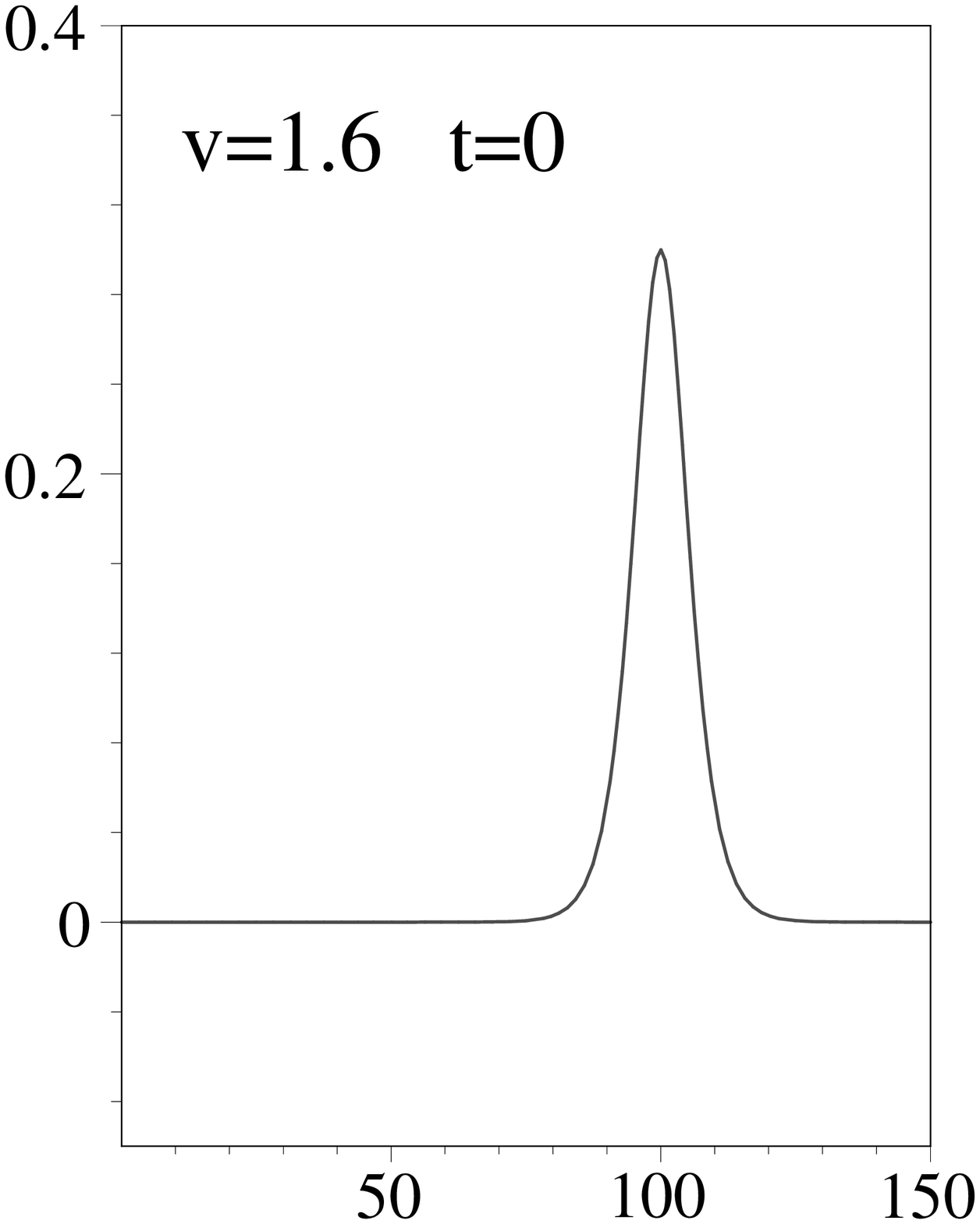,height=4cm,width=4cm}
\epsfig{file=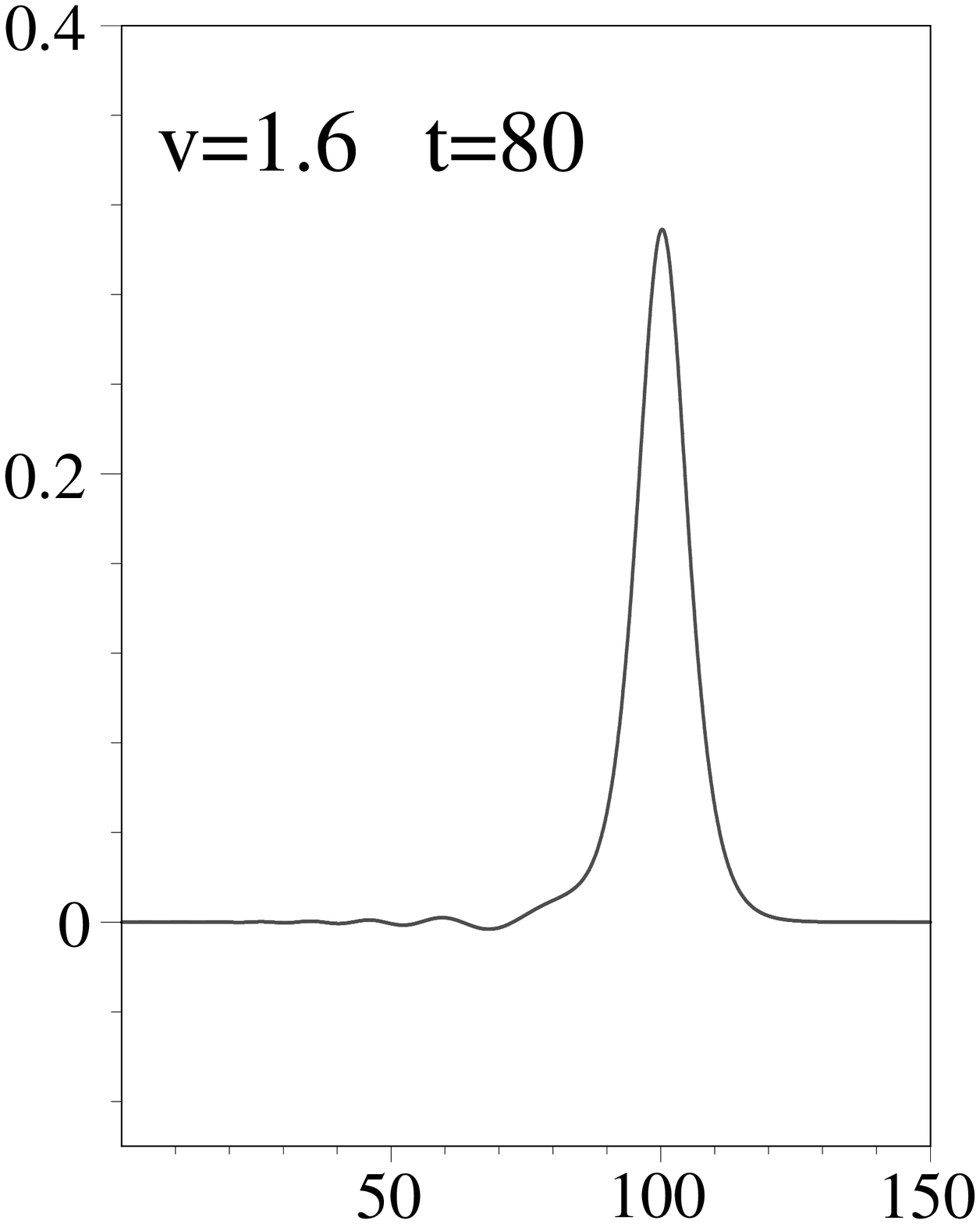,height=4cm,width=4cm}
\epsfig{file=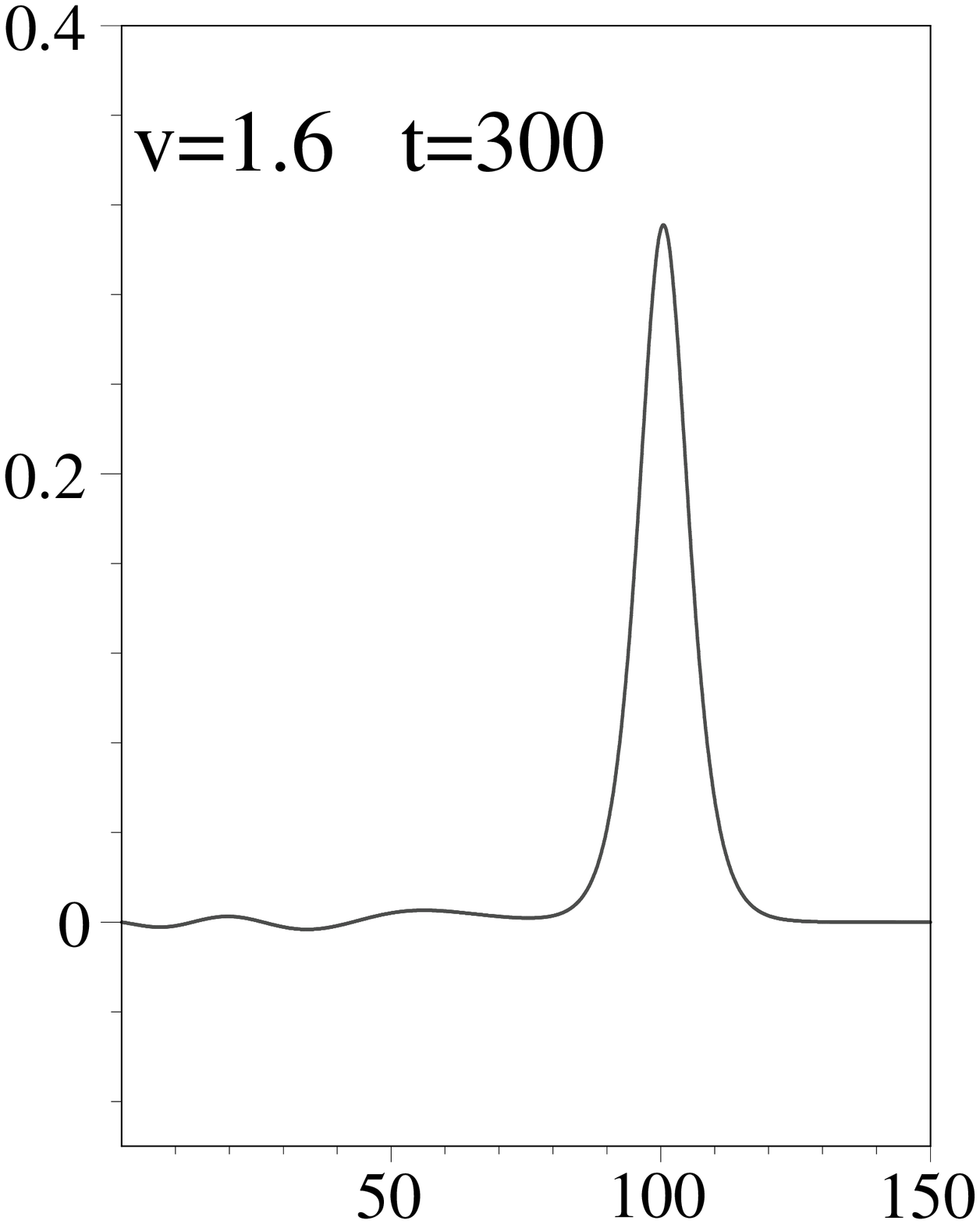,height=4cm,width=4cm}}
\centerline { \epsfig{file=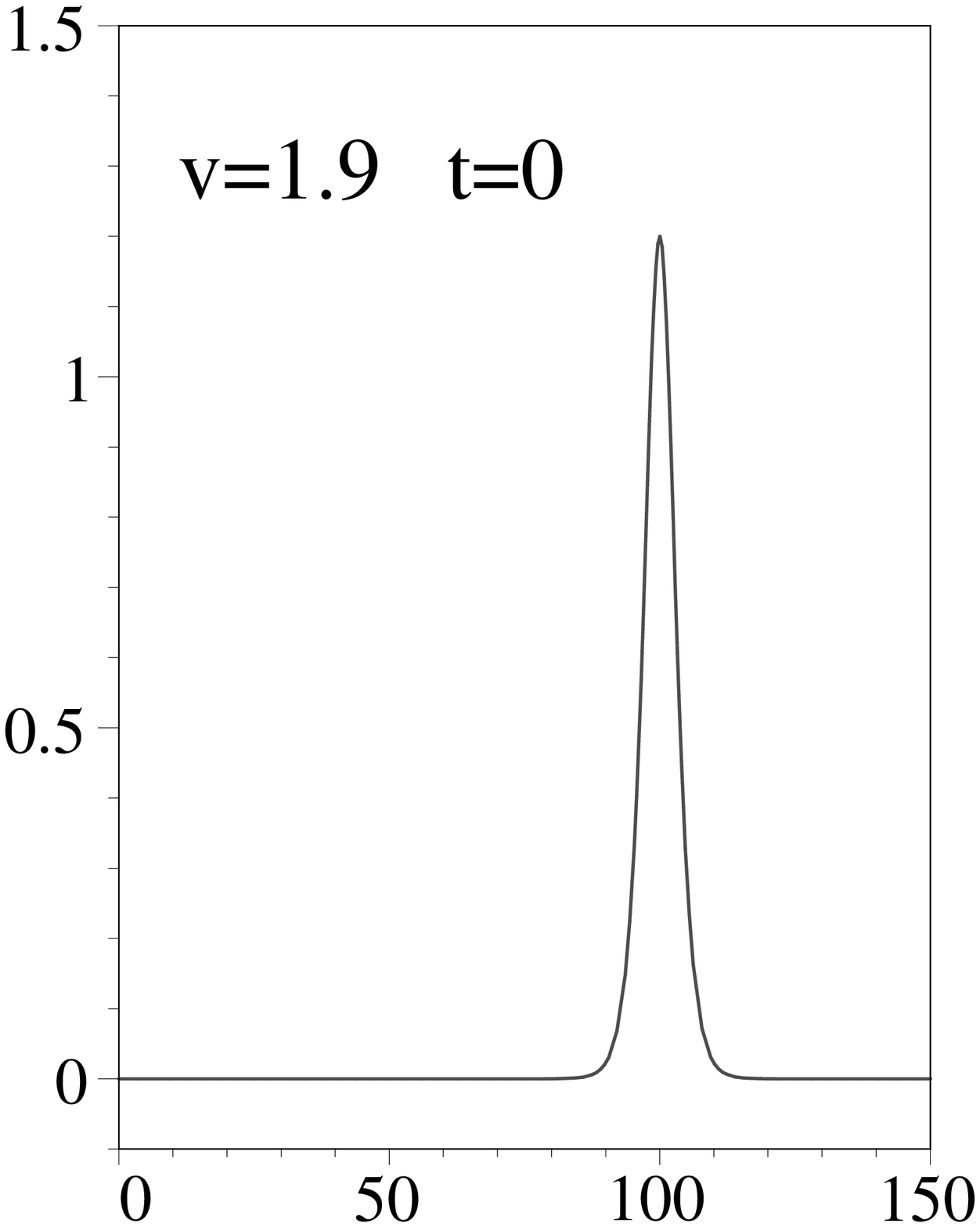,height=4cm,width=4cm}
\epsfig{file=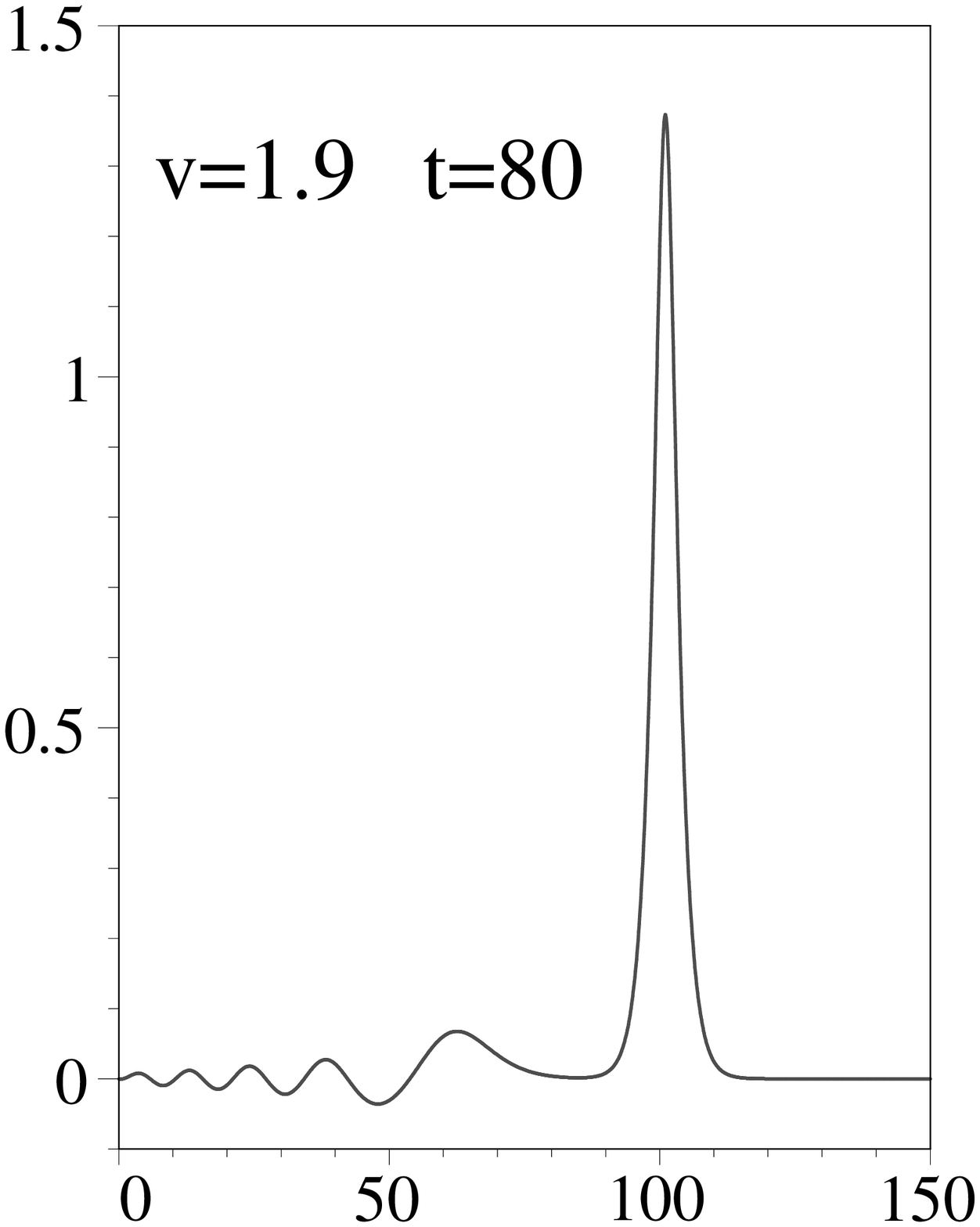,height=4cm,width=4cm}
\epsfig{file=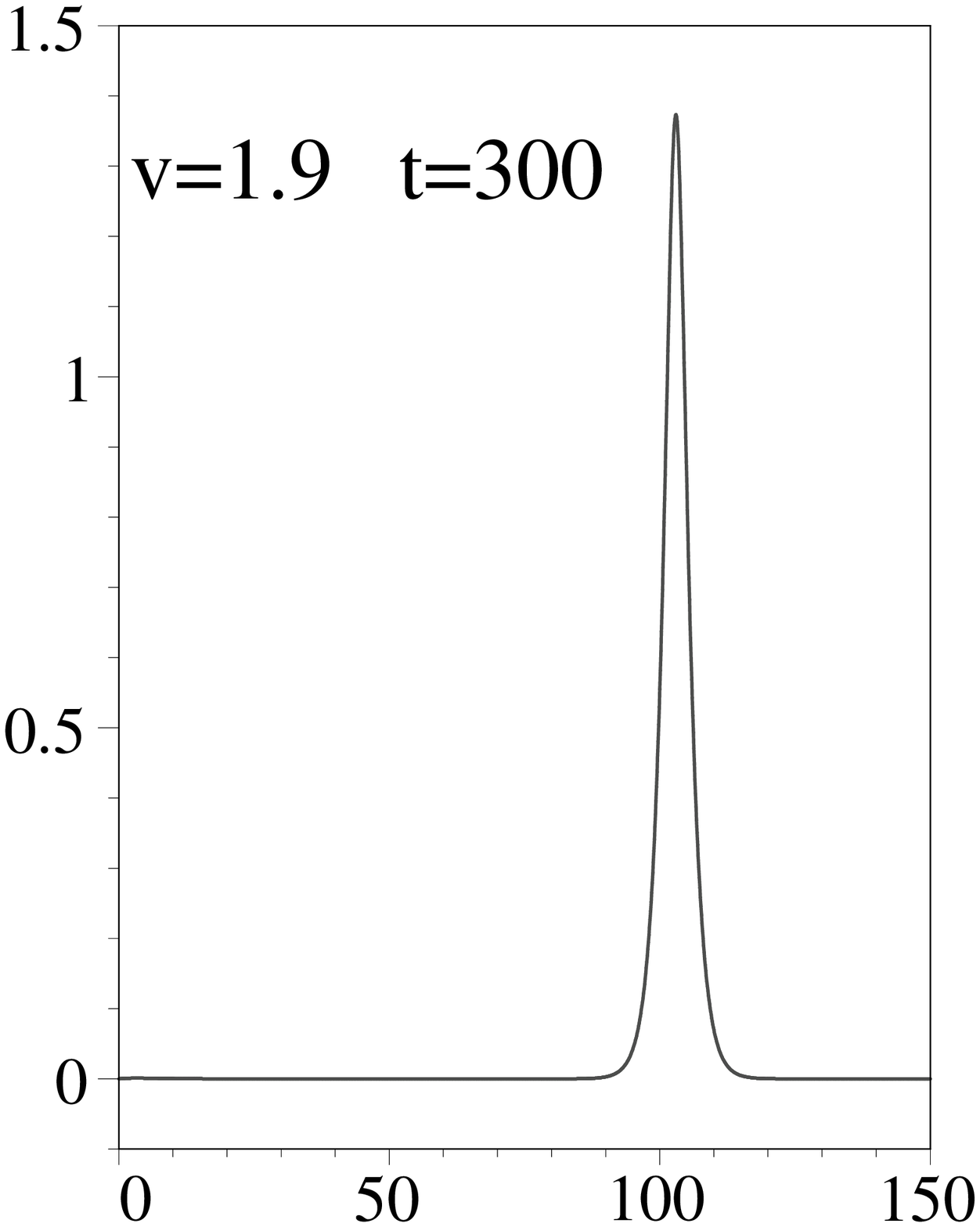,height=4cm,width=4cm}}
\caption {Solitary wave profiles in two examples with $\delta=0.1$ (up)
and $\delta=0.4$ (down) in the rest frame of the approximate solution
\eqref{soliton}. }
\label{fig:solitons}\end{figure}

A few typical examples are displayed in fig.\ref{fig:solitons} where we compare
the solitary wave to the initial condition. As expected, when $\delta<<1$
(i.e. $v\sim 1.5$) the solitary wave resembles much the initial condition while
for larger $\delta$, a pulse reshaping occurs by means of emission of waves
(at phase velocity 1). After reshaping, the solution is remarkably stable.
When $\delta\to 1$ (i.e. $v\to 5/2$), the solitary wave tends to the peakon
solution \eqref{peakon}. 

\begin{figure}[ht]
\centerline {
\epsfig{file=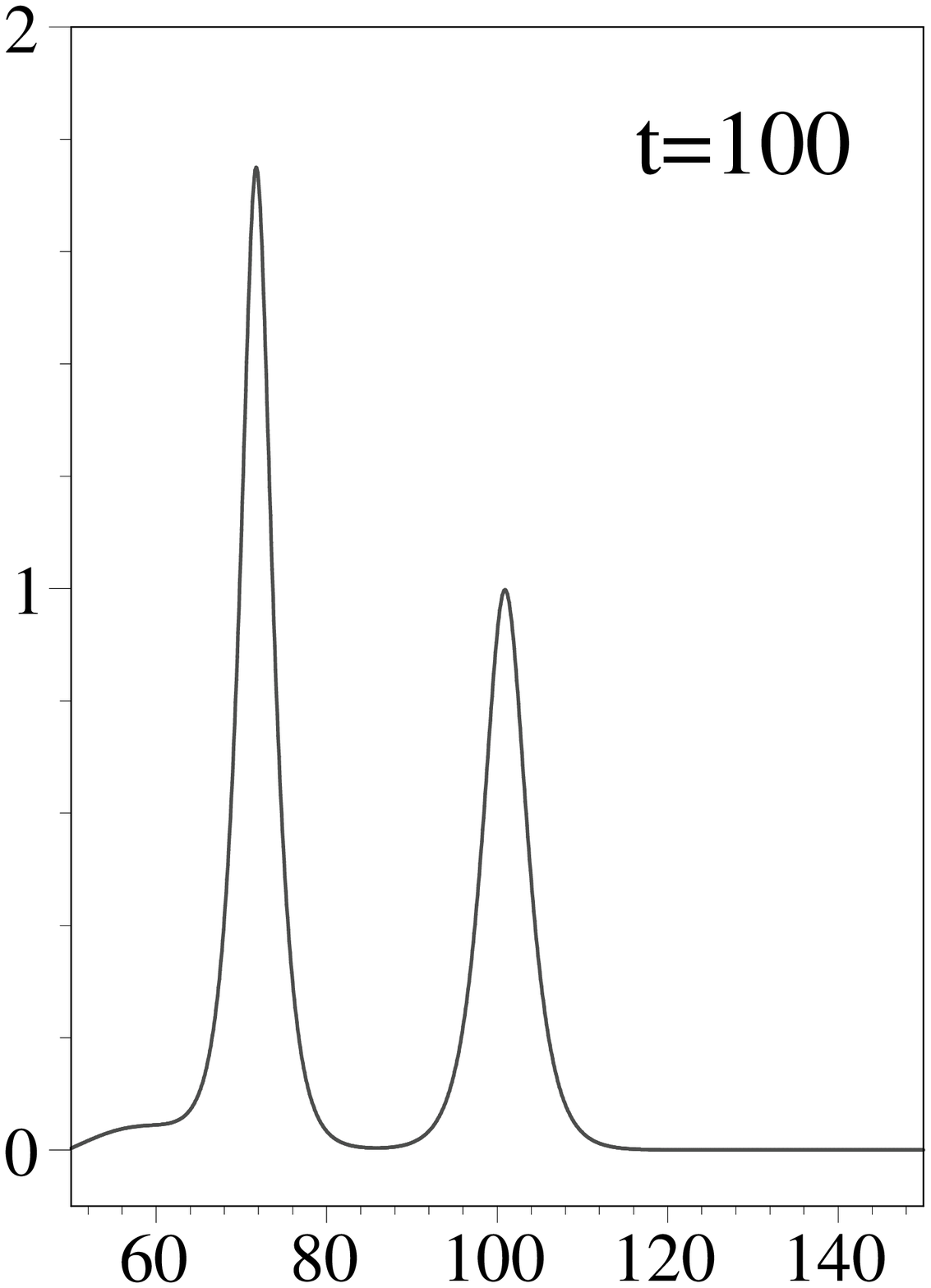,height=3cm,width=4cm}
\epsfig{file=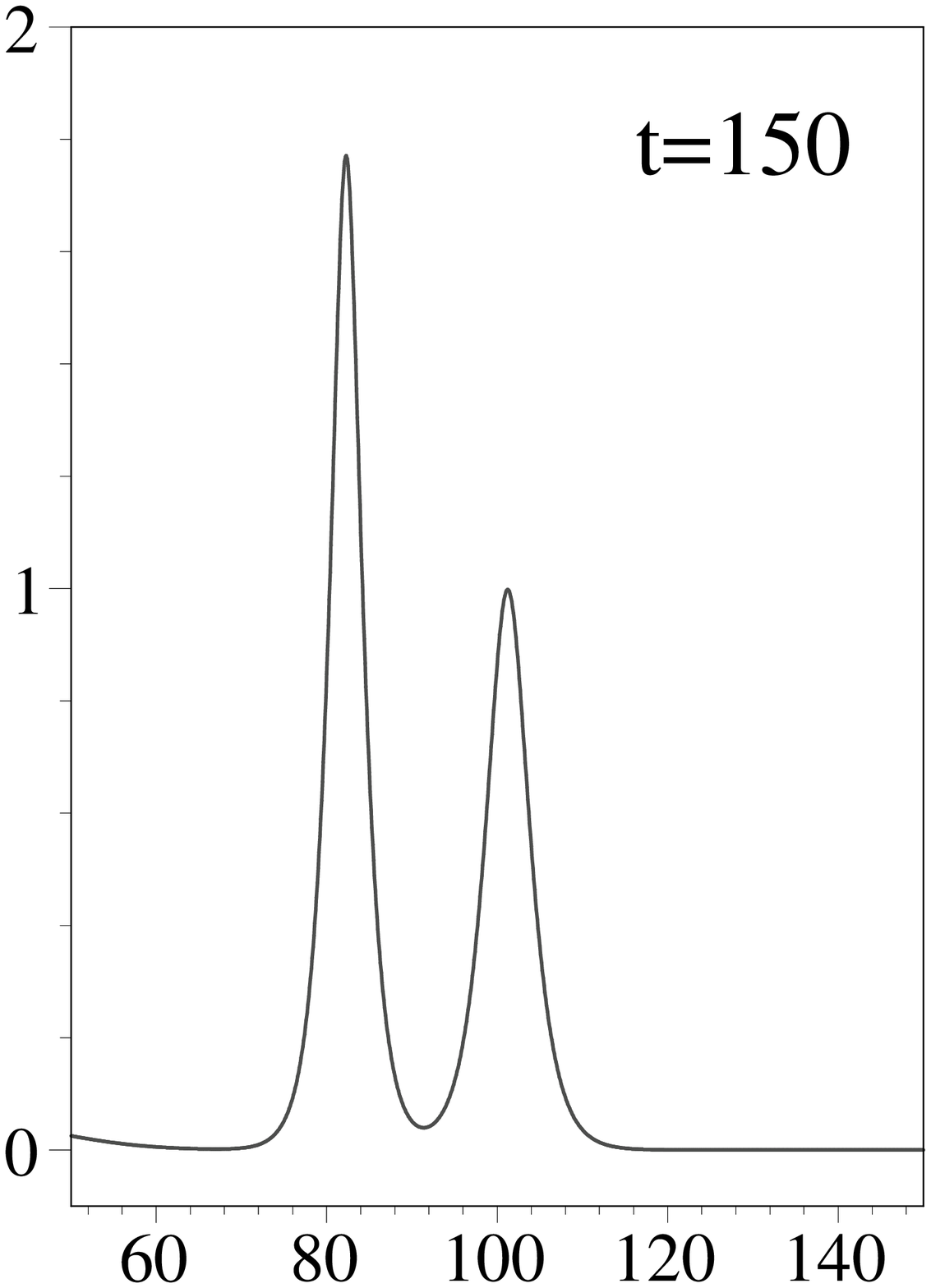,height=3cm,width=4cm}
\epsfig{file=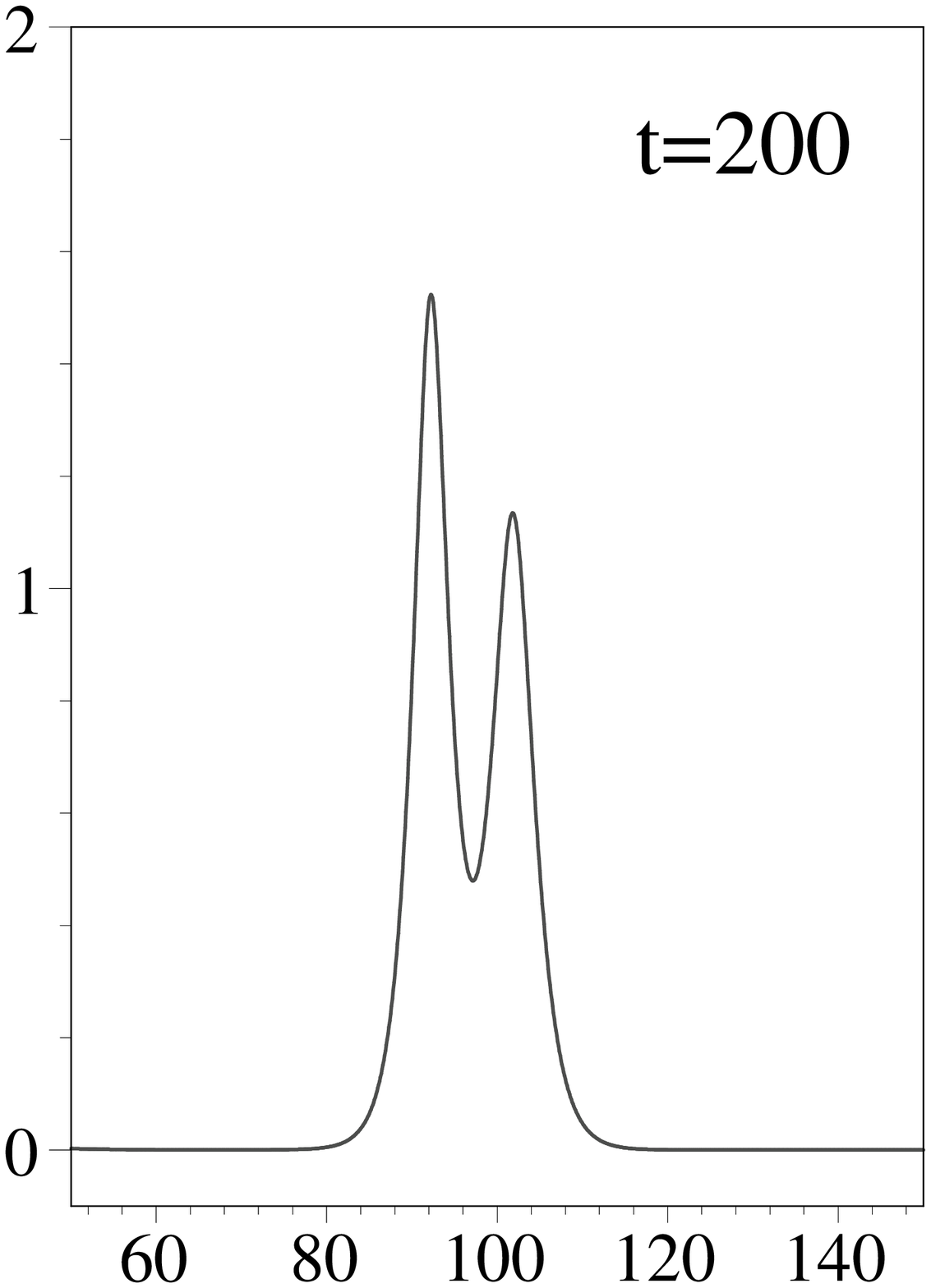,height=3cm,width=4cm}}
\centerline {
\epsfig{file=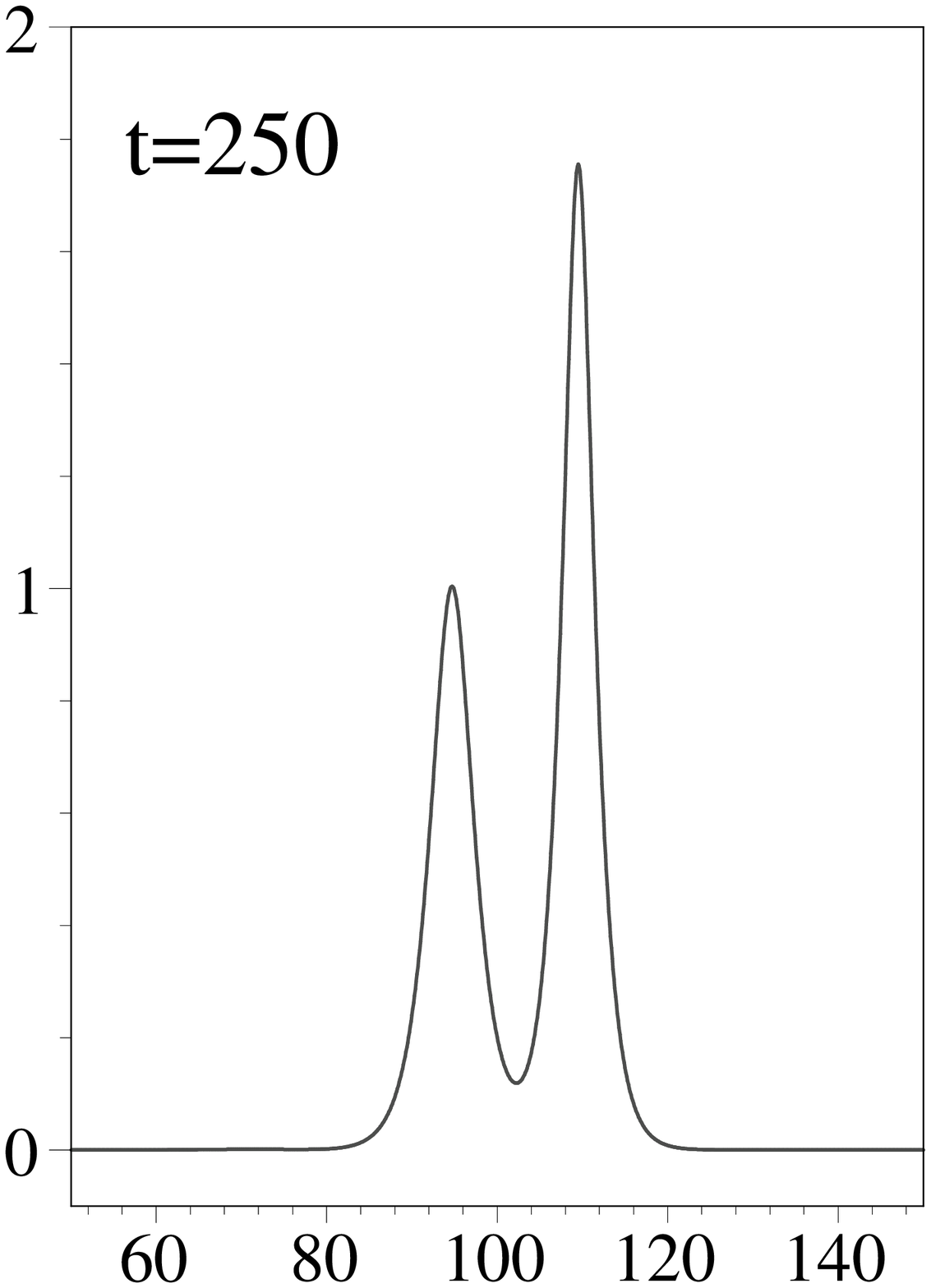,height=3cm,width=4cm}
\epsfig{file=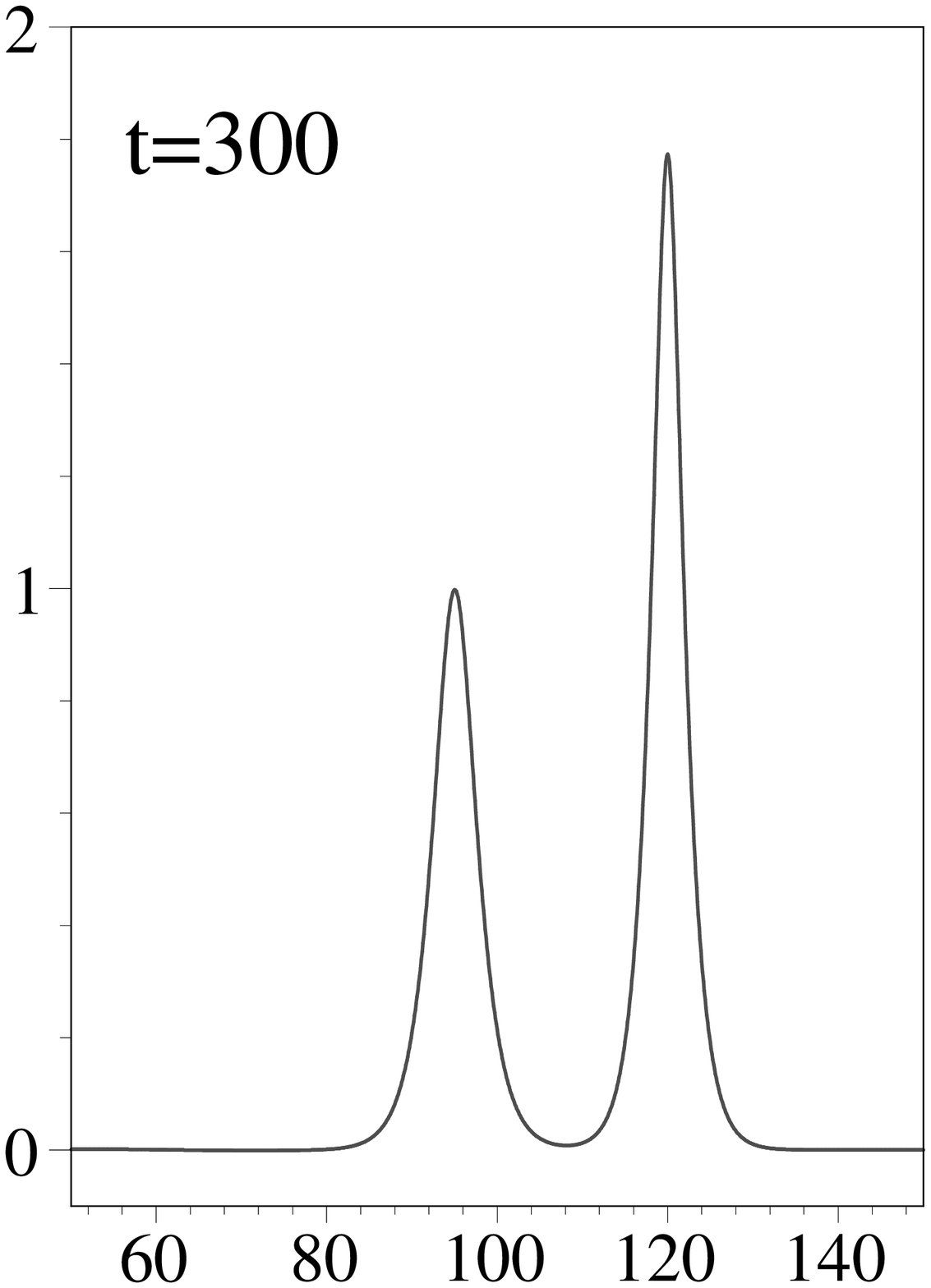,height=3cm,width=4cm}
\epsfig{file=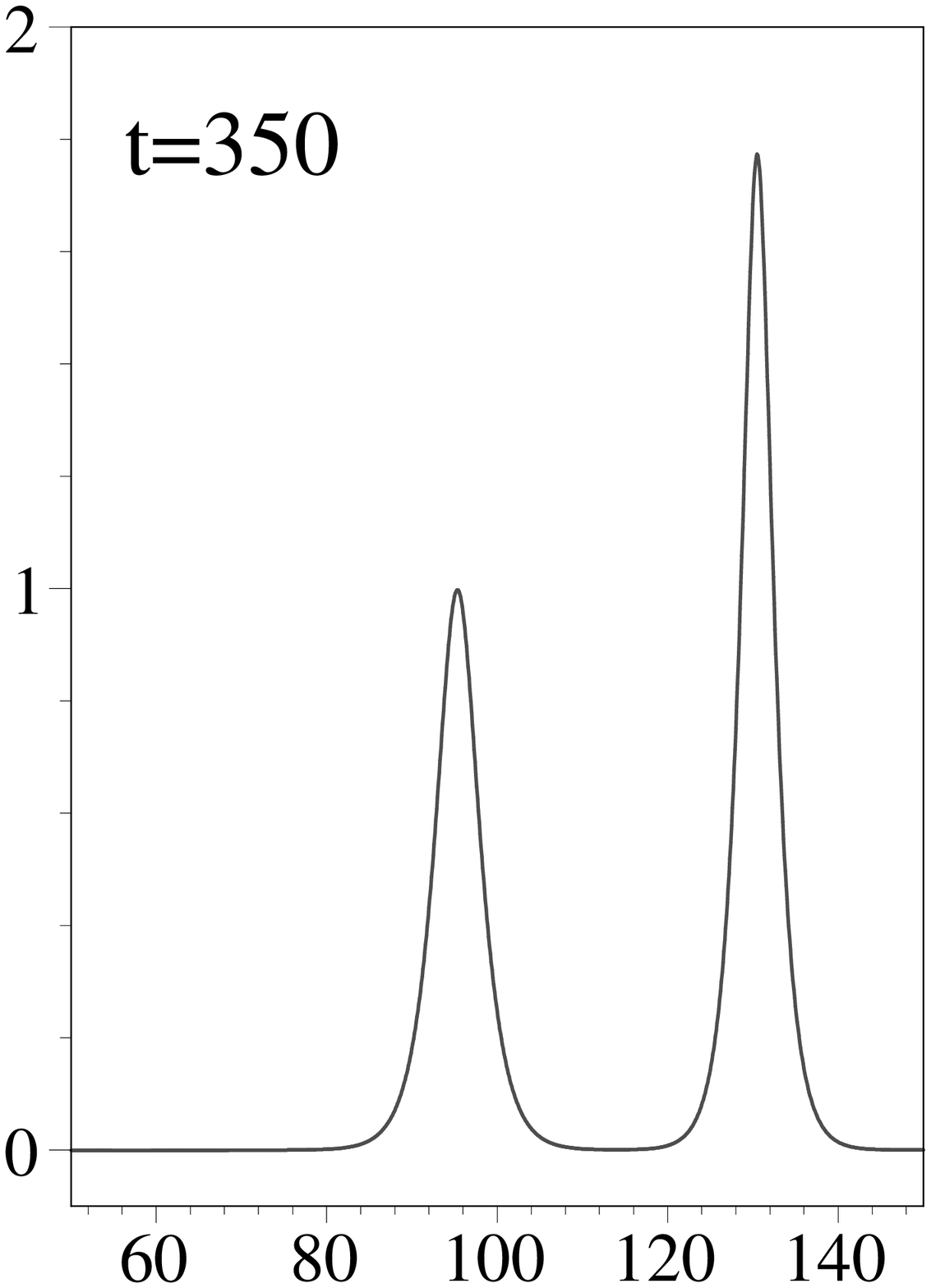,height=3cm,width=4cm}}
\caption {Solitary wave interaction in the rest frame of the soliton at velocity
$1.8$, the other soliton having velocity $2$.}
\label{fig:inter-sol}\end{figure}

The found solitary wave solutions behave like solitons do in an integrable
system. Indeed, after reshaping, a two-soliton interaction effects only the
position of the pulses and does not alter the individual shapes. An example is
diplayed in figure \ref{fig:inter-sol}

\section*{Conclusion and comments}

The asumed  exponential vertical velocity profile has led to a simple model for
surface waves on deep water. The derivation is rigorously made by perturbative
analysis assuming waves with small aspect ratio parameter, which constitutes a
novel approach of the problem. 

The resulting model \eqref{eta} sustains periodic waves which tend to peak as
their amplitude increase and reach a threshold amplitude $\eta_s$ given by
\eqref{threshold}. At this value, the crest angle can be explicitely computed
in terms of the velocity by \eqref{angle}. However, in the region of the peak,
higher order derivatives diverge and the perturbative expansion does not hold
anymore. Hence the precise value of the crest angle should be considered with
care and the peaking only gives an indication of a behavior due to
nonlinearity. 

As mentionned in the introduction,  it is necessary to release the {\it
potential flow} approach which is not consistent with the anzatz
\eqref{anzatz1}. Indeed, assuming the existence of a potential flow $\phi$ by
\begin{equation}(U,W)=\nabla\phi,\end{equation}
the expressions \eqref{anzatz1} and \eqref{anzatz2}, together with
\eqref{cont2}, readily provide
\begin{equation}\phi(x,z,t)=\varphi(x,t)e^z,\quad \varphi_{xx}+\varphi=0.
\end{equation}
Although one may obtain nontrivial time-dependence from the Euler system
in this particular context, the above  linear $x$-dependence of $\varphi(x,t)$ 
cannot catch nonlinear deformations in a consistent way.

Last but not least, it is remarkable that equation (\ref{eqeta}) possess
$k$-dependent coefficients, where $k$ is the wave number of the selected
particular wave. It is something new for  a model describing a wave profile.
Indeed asymptotic non-linear and dispersive long scales models for surface
water waves (as Korteweg-deVries, modified KdV, 
Benjamin-Bona-Mahoney-Peregrine, Camassa-Holm)  have $k$-independent
coefficients. Conversely, $k$-dependent coefficients are common in the context
of modulation of wave trains, as in nonlinear Schr\"odinger, modified NLS,
Davey-Stewartson, etc...

\end{document}